\documentclass{article}

\usepackage{microtype}
\usepackage{graphicx}
\usepackage{subfigure}
\usepackage{booktabs} 

\usepackage{amsmath}

\usepackage{hyperref}
\usepackage{multirow}

\usepackage[accepted]{mlsys2024}

\newcommand{\ours}{vMCU}

\mlsystitlerunning{\ours: Coordinated Memory Management and Kernel Optimization for DNN Inference on MCUs}

\begin{document}

\twocolumn[
\mlsystitle{\ours: Coordinated Memory Management and Kernel Optimization for DNN Inference on MCUs}

\mlsyssetsymbol{intern}{*}
\mlsyssetsymbol{equal}{\dag}

\begin{mlsysauthorlist}
\mlsysauthor{Size Zheng}{pku,uw,equal}
\mlsysauthor{Renze Chen}{pku,equal}
\mlsysauthor{Meng Li}{pku}
\mlsysauthor{Zihao Ye}{uw}
\mlsysauthor{Luis Ceze}{uw,octoai}
\mlsysauthor{Yun Liang}{pku}
\end{mlsysauthorlist}

\mlsysaffiliation{pku}{School of Computer Science, Peking University, Beijing, China}
\mlsysaffiliation{uw}{School of Computer Science \& Engineering, University of Washington, Seattle, United States}
\mlsysaffiliation{octoai}{OctoAI}

\mlsyscorrespondingauthor{Yun Liang}{ericlyun@pku.edu.cn}

\mlsyskeywords{Microconroller, Memory Optimization, AI}

\vskip 0.3in



\begin{abstract}


IoT devices based on microcontroller units (MCU) provide ultra-low power consumption and ubiquitous computation for near-sensor deep learning models (DNN).
However, the memory of MCU is usually 2-3 orders of magnitude smaller than mobile devices, which makes it challenging to map DNNs onto MCUs.
Previous work separates memory management and kernel implementation for MCU and relies on coarse-grained memory management techniques such as inplace update to reduce memory consumption.

In this paper, we propose to coordinate memory management and kernel optimization for DNN inference on MCUs to enable fine-grained memory management. The key idea is to virtualize the limited memory of MCU as a large memory pool. Each kernel divides the memory pool into kernel-specific segments and handles segment load and store while computing DNN layers.
Memory consumption can be reduced because using the fine-grained segment-level memory control, we can overlap the memory footprint of different tensors without the need to materialize them at the same time. 
Following this idea, we implement \ours{} for DNN inference on MCU.
Evaluation for single layers on ARM Cortex-M4 and Cortex-M7 processors shows that \ours{} can reduce from $12.0\%$ to $49.5\%$ RAM usage and from $20.6\%$ to $53.0\%$ energy consumption compared to state-of-the-art work. For full DNN evaluation, \ours{} can reduce the memory bottleneck by $61.5\%$, enabling more models to be deployed on low-end MCUs.
\end{abstract}

]

\printAffiliationsAndNotice{\mlsysEqual}

\section{Introduction}


IoT devices based on always-on microconroller units (MCU) have been widely used in signal processing, environment monitoring, and robotics. Recently, there has been an increasing trend to deploy deep learning models (DNN) on IoT devices~\cite{mcunet, tinyml, bin-net, train-sparse, jingtonghu-dac}, which has enabled various applications including personalized healthcare, unmanned retail, and autopilot. However, it remains challenging to map the DNN layers to MCU because the MCU has very limited compute and memory resources. For example, a STM32-F411RE SoC~\cite{stm32-f411re} with ARM Cortex-M4 processor only has $128$KB RAM, which is 4-5 orders of magnitudes smaller than that of server-class accelerators (e.g., A100 GPU~\cite{a100}) and even 2-3 orders of magnitudes smaller than that of mobile devices (e.g., Kirin-990 SoC~\cite{kirin990}). 
By contrast, even one single convolution layer (image size $=56\times 56$, input/channel $=64$, Int8 precision) from ResNet-50~\cite{resnet} requires around $401.4$KB RAM storage.
This presents huge challenges to model deployment on MCU.

Redesigning DNNs using neural architecture search (NAS)~\cite{proxyless-nas} makes it possible to deploy DNNs to MCUs. NAS can help retrain a new network with much less parameters according to hardware-specific configurations (memory size, compute latency, etc.), but the solution is inevitable hardware-specific. For example, MCUNet-320KB-ImageNet, a typical network proposed by MCUNet~\cite{mcunet, mcunetv2}, can only be deployed to MCUs with memory larger than 320KB.
For smaller devices such as STM32-F411RE (128KB RAM), this network is not executable.
As a result, there is an urgent need for post-training optimizations to further reduce DNN memory consumption on MCU from a system perspective.


Existing efforts~\cite{mcunet, mcunetv2, tflite-micro, chaos, shuzhang} mainly rely on a coarse-grained tensor-level memory management module to schedule the memory usage of each layer in DNNs. 
The memory management module is decoupled from kernel implementation (each kernel implements the computation of one or many layers) for the ease of development and good maintainability.
For example, TinyEngine~\cite{mcunet, mcunetv2} maintains a memory pool for tensors. Before the execution of each kernel, the input, output, and workspace tensors are allocated in the memory pool and passed to kernel as parameters. To save memory, these tensors may overlap with each other at the granularity of the whole tensor (e.g., for depthwise layers). When full tensor overlapping is infeasible, no memory optimization will be performed (e.g., for fully connected layers).
With this method, kernel execution results are always correct even without awareness of any memory overlapping. So existing frameworks can use libraries such as CMSIS-NN~\cite{cmsis-nn} and CMIX-NN~\cite{cmix-nn} or either implement their own kernels without consideration for memory optimization~\cite{mcunet}.

However, tensor-level memory management misses the chance to achieve the optimal memory footprint.
For most layers in DNNs, full tensor overlapping is infeasible. For example, 2D convolution and fully connected layers can't inplace update their outputs due to the frequent reuse of the input data. These layers become the memory bottleneck for the whole DNN even if we can apply inplace optimizations to other elementwise and depthwise layers.

To address this issue, in this paper, we propose \textbf{\ours} to enable fine-grained memory management.
We virtualize the MCU memory and exploit the chance of partial tensor overlapping in DNN.
In detail, the data elements in input tensors may have different lifetime, we group elements into segments and set the lifetime of the segment as the maximum lifetime of the elements in the segment.
After this, we can schedule memory allocation at the granularity of segments.
When computing layers such as 2D convolution, the segments of output tensors can overlap with the segments of input tensors, enabling memory footprint reduction.

Implementing segment-level memory management is challenging because the kernel computation should be aware of segment overlapping. To address this challenge, we propose to coordinate memory management with kernel optimization.
In detail, for memory management module level, segments are maintained in a circular buffer, while the segment size is dependent on each kernel. The memory management module is responsible for input, output, and workspace tensor allocation in the circular buffer.
For kernel optimization level, each kernel first chooses a kernel-specific segment size and get input/output/workspace tensors from memory management module. Then, two-level tiling is applied to each kernel: the outer level coordinates with segment size, while the inner level coordinates with compute instruction lane size.
Finally, segment overlapping and replacement is done for the outer level tiles. When segment pointers are out of boundary, they will be reset in the circular buffer using address modular operations.

To reduce the development and maintaining difficulty, we further provide a Python interface for kernel implementation with reusable intrinsic and code generation support for MCU.

In summary, we make the following contributions:
\begin{enumerate}
    \item We propose to coordinate the memory management module and kernel optimization to enable segment-level memory overlapping, which effectively reduce memory footprint for DNN.
    \item We design a segment-level memory management technique and implement a kernel library for MCU that supports partial memory overlapping for both single layer and multiple layers.
    \item We effectively reduce the memory footprint of both single layer and whole network on real MCUs without any retraining overhead.
\end{enumerate}
A good property of \ours{} is that it can reduce the memory footprint of linear structure DNNs, which is infeasible in previous work~\cite{chaos, hierarchy}.
Evaluation for single layer on ARM Cortex-M4 and Cortex-M7 processors shows that \ours{} can reduce from $12.0\%$ to $49.5\%$ RAM usage and from $20.6\%$ to $53.0\%$ energy consumption compared to state-of-the-art work. For end-to-end linear structure DNN evaluation, \ours{} can reduce the memory bottleneck by $61.5\%$. 

\section{Background and Motivation}

\begin{figure*}[!t]
\centering
\includegraphics[width=\textwidth]{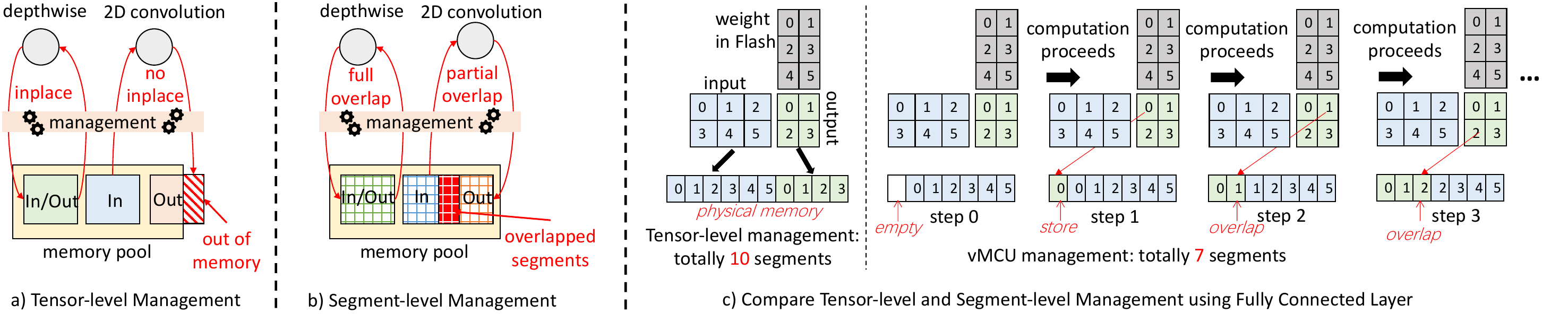}
\DeclareGraphicsExtensions.
\caption{a) and b): Compare Tensor-level memory management and segment-level memory management. c): Motivational example.
}
\label{fig:motivation}
\end{figure*}

\begin{table}[t]
  \centering
  \footnotesize
  \caption{Features of accelerators, mobile devices, and MCUs.}
\begin{small}
  \begin{tabular}{c|c|c|c}
  \hline
    \textbf{Hardware} & \textbf{Memory} & \textbf{Storage} & \textbf{SW Support}\\ 
  \hline
    \textbf{A100} & 40GB & TB-PB  & CUDA runtime\\
    
  \hline
    \textbf{Kirin-990} & ~8GB & ~256GB & OS (Linux)\\
    
  \hline
    \textbf{F411RE} & 128KB & 512KB & None\\
  \hline
  \end{tabular}
  \label{table:compare}
\end{small}
\end{table}

\subsection{Architecture Features of Microcontrollers}
Microcontroller units (MCU) provide ultra-low energy consumption and cheap computation for near-sensor processing.
We summarize the memory comparison among cloud accelerators, mobile devices, and MCUs in Table~\ref{table:compare}.
For cloud accelerators, we use Nvidia A100 GPU as example; for mobile devices, we use Kirin-990 SoC (with ARM Cortex-A7 processors) as example; and for MCUs, we use STM32-F411RE (with ARM Cortex-M4 processor). MCUs are often equipped with limited memory and storage resources (KB-MB) and their memory hierarchies are specially simplified to meet the low-power and always-on requirements. The DNN programs cannot run if the footprint exceeds the capacity.
There is no hardware cache or software operating system on MCUs, and the data mapping can be only handled statically in programs (by allocating data arrays and operating directly through pointers).
These features make it hard to deploy DNNs to MCU. 
When deploying DNN models on MCUs, the major challenge is to map the enormous amount of tensor data to the limited memory resource without much loss of performance.

\subsection{Reduce DNN Size With NAS}
The most effective approach to reduce the size of DNNs for MCUs is NAS~\cite{proxyless-nas, mcunet, mcunetv2}.
NAS trains a super network and aims to find the optimal sub-network of the super network that both achieves good accuracy and conforms to the hardware constraints (e.g., memory resource and latency).
For example,
MCUNet~\cite{mcunet, mcunetv2} redesigns a series of CNN models with smaller convolutions.
Usually, the more computing operations in a network, the higher accuracy it can achieve. As a result, NAS always produces the largest network possible for the target hardware for good accuracy results.
So the result networks are hardware-specific and many of them can't be deployed to smaller MCUs due to insufficient memory resources.
For example, MCUNet-320KB-ImageNet~\cite{mcunet} can't be deployed to STM32-F411RE SoC that only has 128KB RAM.
This problem is viewed as the \textit{last mile to go} for DNN deployment on MCU, which calls for a post-training solution to further reduce the memory footprint at a system level.

\subsection{Tensor-level Memory Management on MCU}
Existing works~\cite{tflite-micro, cmsis-nn, jason-xue-dac, tvm, mcunet, mcunetv2, chaos, hierarchy} focus on tensor-level memory management on MCU.
The input, output, and workspace tensors for each kernel are maintained in a memory pool.
To save memory, the tensors may be overlapped.
For example, depthwise convolution allows input and output overlapping because the inner computation has no inter-channel data reuse.
The tensor-level management makes it possible to decouple memory management from kernel implementation.
For example, TensorFlow Lite Micro~\cite{tflite-micro} uses kernels from CMSIS-NN~\cite{cmsis-nn} and schdules the tensors for these kernels during code generation; Serenity~\cite{chaos} uses dynamic programming to find memory optimal execution order for different kernels so that some short-lifetime tensors can be freed and reused for later tensors; HMCOS~\cite{hierarchy} is similar to Serenity but improves the memory management for local sub-graph optimization.

The limitation of tensor-level memory management is that, when full tensor overlapping is infeasible, no memory optimization will be performed. For example, as shown in Figure~\ref{fig:motivation} part a), for a 2D convolution, its input and output tensor can't be fully overlapped, so the management module allocates different memory spaces for them. For small devices with insufficient memory resources, out of memory error will occur.
As we will point out in the next section, we can still exploit memory overlapping to reduce memory footprint for such scenarios.

\subsection{Motivational Example}
We use the example in Figure~\ref{fig:motivation} part c) to explain how to reduce memory footprint by partially overlapping tensors.
This example compares how tensor-level memory management and our proposed segment-level memory management differ in allocating memory for inputs and outputs for a fully connected layer.
We assume the fully connected layer input tensor can be divided into $2\times 3$ segments and output tensor can be divided into $2\times 2$ segments.
The weight tensor is constant and is placed in Flash memory, so we don't consider it in memory management.
Fully connected layer is not suitable for inplace update, so tensor-level memory management allocates two different memory spaces for input and output tensors, requring totally 10 segments.
On the contrary, segment-level management only needs 7 segments to complete the computation.
Initially, only input tensor (6 segments) exist in memory.
The first segment of output occupies an empty segment in memory.
Start from the second step, the input tensor fragment at the front is freed and a newly updated output segment is put to the same place in memory.
After five steps, all the output segments are produced correctly and the computation of the fully connected layer ends.

One important factor in memory management is the number of empty segments allocated for output tensor. If we don't allocate enough segments for output tensor, the newly updated output segments will incorrectly replace the segments of input tensor, causing silent error in correctness. If we allocate too many empty segments, then we will waste memory space and get suboptimal memory footprint.

The opportunity of further reducing memory footprint using segment-level memory management motivates us to design \ours{}.
In the following sections, we present the details of \ours{}.

\section{Overview of \ours{}}
\label{sec:overview}

\begin{figure}[!t]
\centering
\includegraphics[width=3.3in]{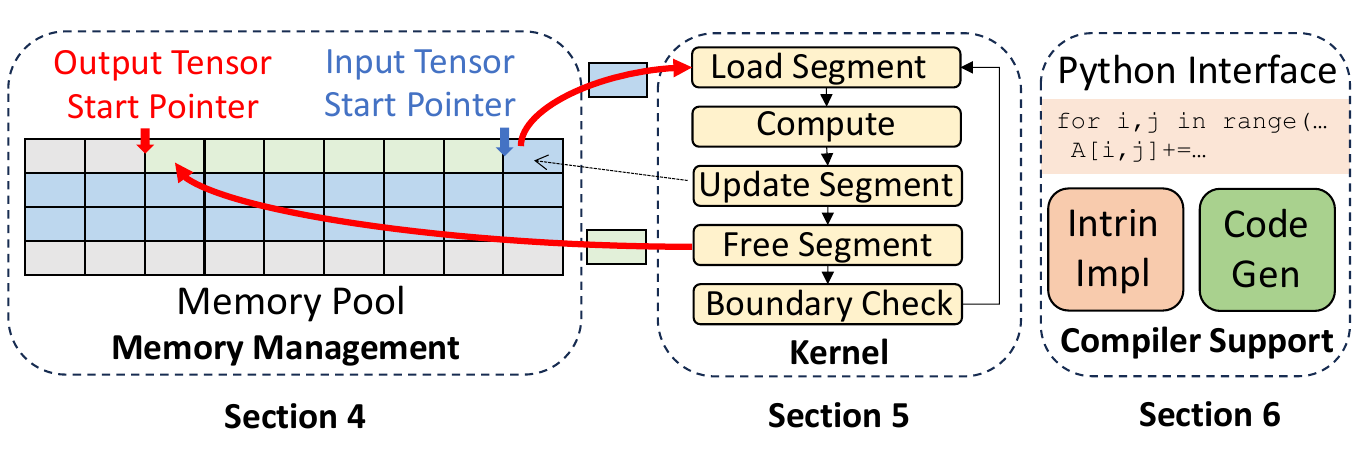}
\DeclareGraphicsExtensions.
\caption{Overview of \ours{}.
}
\label{fig:overview}
\end{figure}

The overview of \ours{} is shown in Figure~\ref{fig:overview}. 
\ours{} is composed of three parts: memory management part, kernel design part, and compiler support part.
Memory management part maintains a memory pool of segments. Before the execution of each kernel, this part determines the start pointers for input and output tensors according to kernel-specific segment size as well as the segment arrangement order in memory. This part is explained in Section~\ref{sec:algorithm}.
Kernel design part provide kernel design methods for DNN layers, including both single layer and multiple layers. We mainly focus on 2D convolution and fully connected layers for single layer and focus on inverted bottleneck module for multiple layers because they account for the most layers in DNNs on MCUs. 
Overall, one kernel is composed of five steps: segment loading, computation on the segment, update the output segment to memory, free a used segment, and boundary check for possible pointer out of boundary.
The details are explained in Section~\ref{sec:library}.
Compiler support part is designed to reduce the development difficulty with \ours{}. We provide a Python programming interface for kernel implementation in \ours{}. The compiler for \ours{} is composed of two parts: intrinsic and code generation support, which is explained in Section~\ref{sec:compiler}.

\section{Segment-level Memory Management}
\label{sec:algorithm}

In this section, we explain how the segment-level memory management part setups the initial input/output tensor pointers in memory.

Before we formulate the problem, we need to clarify several details.
First, the order of segment arrangement in memory is critical to kernel computation correctness.
We assume the segments of different tensors are stored in memory in row-major order.
For example, as shown in the example in Figure~\ref{fig:motivation}, the segments of input and output tensors are in row-major order, which aligns well with the computation order in kernel, guaranteeing the correctness of output data.
For cases where column-major order is needed, the memory management method should be similar, so we don't discuss it for simplicity.
Second, we mainly consider dense workloads because most DNNs on MCUs use dense tensors with quantization~\cite{mcunet, mcunetv2, cmsis-nn, tflite-micro}. So we can safely assume all the segments are contiguous in memory.


To formulate the memory management problem, we first formally define the memory pool with segments.
For a given segment size $Seg$ bytes (determined by kernel implementation, a tunable hyperparameter) and a given memory size $MemCap$ bytes of MCU, a memory pool is a circular buffer array

\begin{small}
\begin{equation*}
\begin{aligned}
 & Pool[\frac{MemCap}{Seg}], \\
 \text{where} \ Pool[i] \ & \text{is a vector of}\ Seg \ \text{bytes}, \ 0 \le i < \frac{MemCap}{Seg}
\end{aligned}
\end{equation*}
\end{small}
$Pool$ is circular, which means when accessing $Pool$, the address will be calculated using a modulo operation:

\begin{small}
\begin{equation*}
\begin{aligned}
Pool[addr] = Pool[addr\ \%\ \frac{MemCap}{Seg}].
\end{aligned}
\end{equation*}
\end{small}


Considering that kernel computation is done in the granularity of segment, we can formulate the iteration domain of a kernel as a set composed of iteration instances $S[\vec{i}]$:

\begin{small}
\begin{equation*}
\begin{aligned}
\{S[\vec{i}] : \mathbf{H} \vec{i} + \vec{B} < 0 \}
\end{aligned}
\end{equation*}
\end{small}where $\mathbf{H}$ is a matrix and $\vec{B}$ is a vector. They are used to represent the iteration boundaries as affine constraints. $\vec{i}$ is a vector of iteration variables.
Each iteration instance $S[\vec{i}]$ corresponds to one computation step on a segment.

Each iteration instance $S[\vec{i}]$ accesses its input/output tensor data $T$ in the unit of segments, which can be also represented in affine transformations, we call this access function:
\begin{small}
\begin{equation*}
\begin{aligned}
\{S[\vec{i}] \rightarrow T[\vec{u}] : \vec{u} = \mathbf{A_u} \vec{i} + \vec{V_{u}} \}
\end{aligned}
\end{equation*}
\end{small}where $\mathbf{A_u}$ is a matrix and $\vec{V_u}$ is a vector. We name them as access matrices.
$\vec{u}$ is a vector of access indices.
In Figure~\ref{fig:gemm-example} we show the iteration domain and access functions of a GEMM example. The iteration domain is shown in part b), and the access functions along with their matrix representations are shown in parts c) and e). There is no $\vec{V_u}$ vector in the access matrices because there is no index offsets in GEMM program in part a).

Using the access functions and the assumption that all segments are arranged in row-major order in memory, we can derive the segment access address for each iteration instance $S[\vec{i}]$:

\begin{small}
\begin{equation*}
\begin{aligned}
\{S[\vec{i}] \rightarrow T[\vec{u}] \rightarrow Pool[addr] : addr = \vec{L}_{addr} \vec{u} + b_{off} \}
\end{aligned}
\end{equation*}
\end{small}where $\vec{L}_{addr}$ is a vector used to map multi-dim indices $\vec{u}$ to linear address according to row-major order, we call it mapping vector; $b_{off}$ is an address offset. 
Using Figure~\ref{fig:gemm-example} as an example, we show its mapping vectors and offsets in part d) and f). The mapping vector of tensor $In$ is $[K, 1]$ because tensor $In$ is of shape $[M, K]$, so its row-major memory strides are $[K, 1]$. Similarly, the mapping vector of tensor $Out$ is $[N, 1]$. Their offsets $b_{In}$ and $b_{Out}$ determine their initial addresses in memory pool. Note that all the addresses and memory accesses in this example are in the unit of segment.

Using our formulation, the distance between the offsets $b_{In}$ and $b_{Out}$ determines how many empty segments we need to allocate for output tensor at the beginning of kernel execution.
We aim to minimize the peak memory usage without damaging the correctness of computation (i.e., without data race between output and input segments), which equals to minimize $b_{In} - b_{Out}$.
This defines the problem formulation for memory management:



\begin{figure}[!t]
\centering
\includegraphics[width=3.3in]{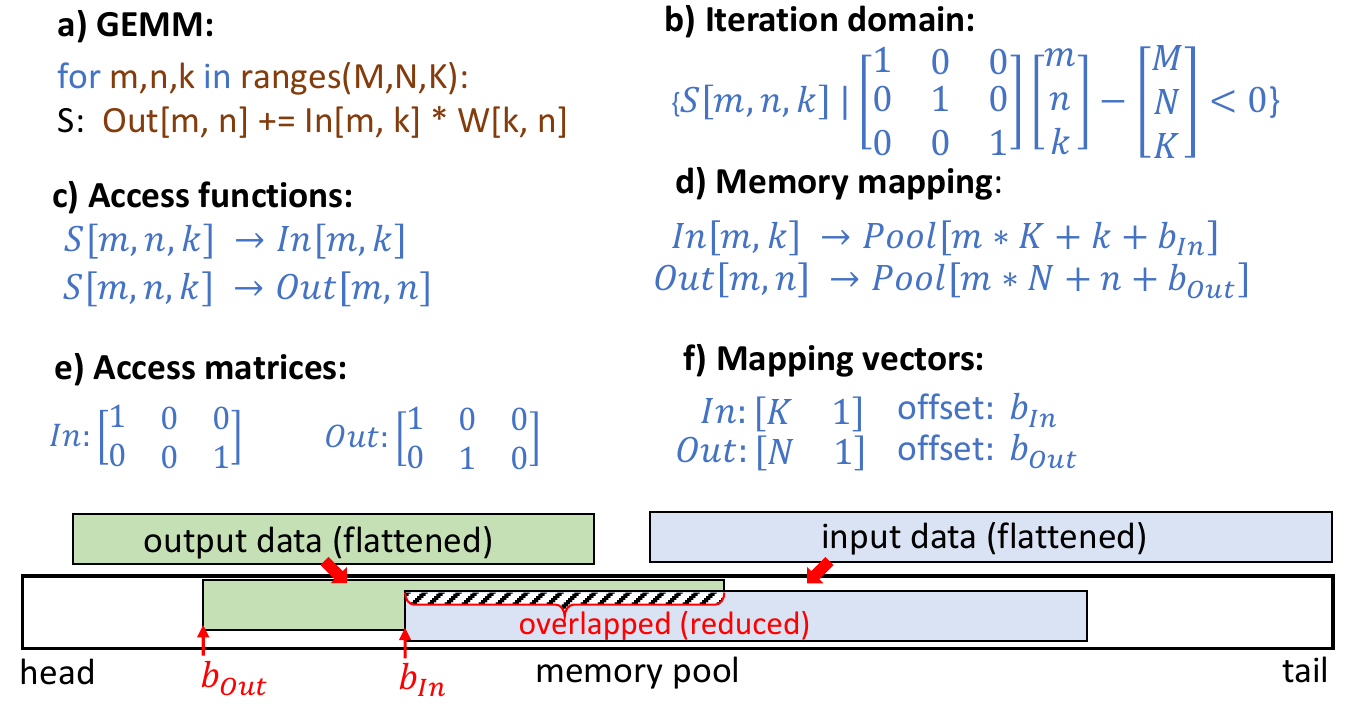}
\DeclareGraphicsExtensions.
\caption{Problem formulation for GEMM example.
}
\label{fig:gemm-example}
\end{figure}


\vspace{-0.3cm}
\begin{small}
\begin{equation}
\label{eq:single-op-formulation}
\begin{aligned}
& \text{min.} \ \ b_{In} - b_{Out}\ \ \text{s.t.}  \ \ \forall{\vec{j} \le \vec{i}}\\
& \vec{L}_{In} (\mathbf{A}_{In} \vec{i} + \vec{V}_{In}) + b_{In} \ge \vec{L}_{Out} (\mathbf{A}_{Out} \vec{j} + \vec{V}_{Out}) + b_{Out}
\end{aligned}
\end{equation}
\end{small}
The constraint in the formulation means that for any iteration instance $\vec{i}$ in the iteration domain, its read address from $In$ is no less than the write address to $Out$ of all the iteration instances before $\vec{i}$ (represented as $\vec{j} \le \vec{i}$), which aligns with the row-major order of segment arrangement and computation.
We can solve the optimization problem by integer linear programming.
Using the solution, we can setup the inital pointers for input and output tensors. The input tensor initial pointer address is determined by the previous layer in DNN (the first layer can choose arbitrary address in memory). The output tensor initial pointer address is set by shifting the input tensor pointer towards the memory pool head by $b_{In} - b_{Out}$ segments.

Using Figure~\ref{fig:gemm-example} as an example, the optimization problem for GEMM is

\begin{small}
\begin{equation*}
\begin{aligned}
& \text{min.} \ \ b_{In} - b_{Out}\\
& \text{s.t.} \ \ (K - N) m - n + k \ge b_{Out} - b_{In},\\
& \forall \ \ 0 \le m < M, \ \ 0 \le n < N, \ \ 0 \le k < K
\end{aligned}
\end{equation*}
\end{small}

The solution to this example is
\begin{small}
\begin{equation*}
\begin{aligned}
\text{MinFootprint} &= max(MN, MK) + b_{In} - b_{out}\\
& = max(MN, MK) + min(N, K) - 1
\end{aligned}
\end{equation*}
\end{small}which means that when $N \le K$, the minimal footprint is $MK + N - 1$; when $N > K$, the minimal footprint is $MN + K - 1$. This result coincides with the result of the example in Figure~\ref{fig:motivation} part c) ($K=3, N=2$), where we only allocate one empty segment ($N-1=1$) to minimize the peak footprint.

\section{Segment-aware Kernel Design}
\label{sec:library}

\subsection{Kernel Design for Single Layer}
\label{sec:library:single}

For a kernel for single layer, a two-level tiling is applied for optimization. The outer level performs data access in unit of segments, while the inner level performs computation on the segments.
Specifically, we explain how to design kernels for 2D convolution and fully connected layers because they account for the most of the layers in DNNs for MCUs.
Although we only discuss two specific layers, the kernel design paradigm is general because the two-level tiling sketch and inner reusable functions are similar for different layers.

\begin{figure}[!t]
\centering
\includegraphics[width=3.3in]{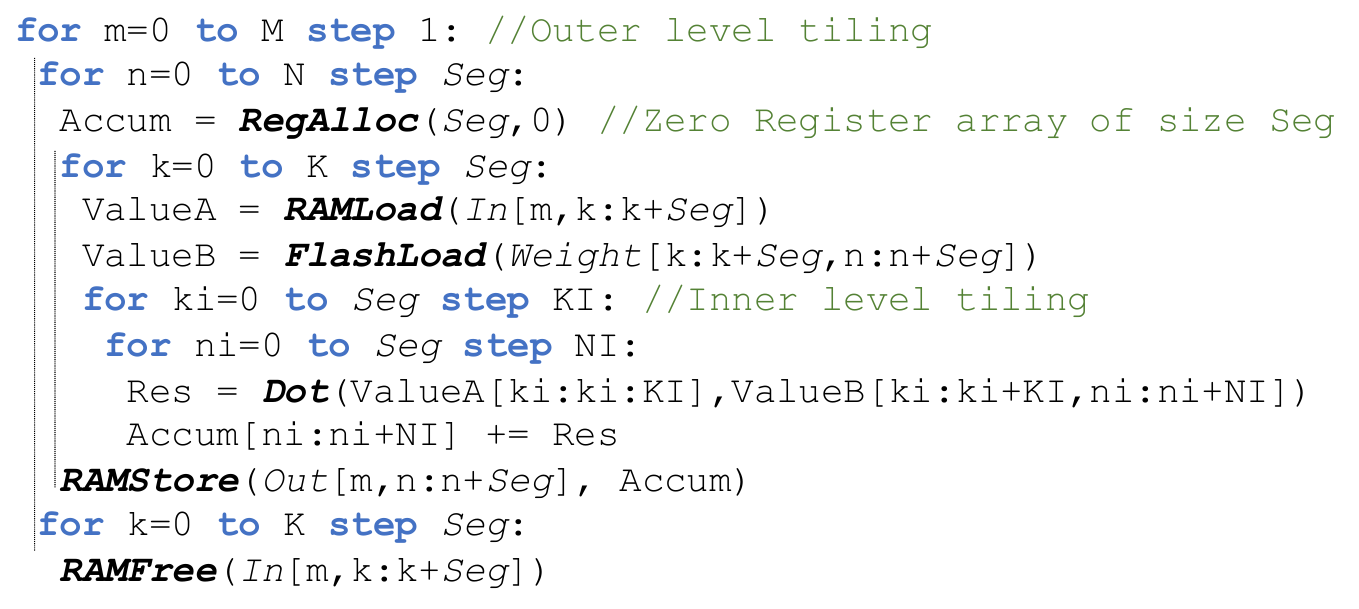}
\DeclareGraphicsExtensions.
\caption{Pseudo code for the kernel of fully connected layer
}
\label{fig:fully-connected}
\end{figure}

We start from the fully connected layer.
For a fully connected layer with input tensor $In[M, K]$, weight tensor $Weight[K, N]$ (in Flash), and output tensor $Out[M, N]$, when the segment size is set as $Seg$, the kernel pseudo code is shown in Figure~\ref{fig:fully-connected}. 
The pseudo code shows a two-level tiling sketch.
The inner level tiling factors $KI, NI$ are hardware-specific parameters, determined by the vector instruction lane length provided by MCU instruction set.
There are five specific functions used in the code: \textit{RegAlloc}, \textit{RAMLoad}, \textit{FlashLoad}, \textit{RAMStore}, and \textit{RAMFree}. As the weight tensor is stored in Flash memory (read-only), we don't consider it in memory management.
\textit{RAMLoad} and \textit{RAMStore} all require boundary check. As we are using circular buffer, the check is done by modulo operation ($addr = addr \% \frac{MemCap}{Seg}$).
As the pseudo code shows, the store of output segments are more frequent than the free of input segments. Allocating enough empty segments in memory pool guarantees that the behavior of this kernel is correct (as explained in Section~\ref{sec:algorithm}).

\begin{figure}[!t]
\centering
\includegraphics[width=3.3in]{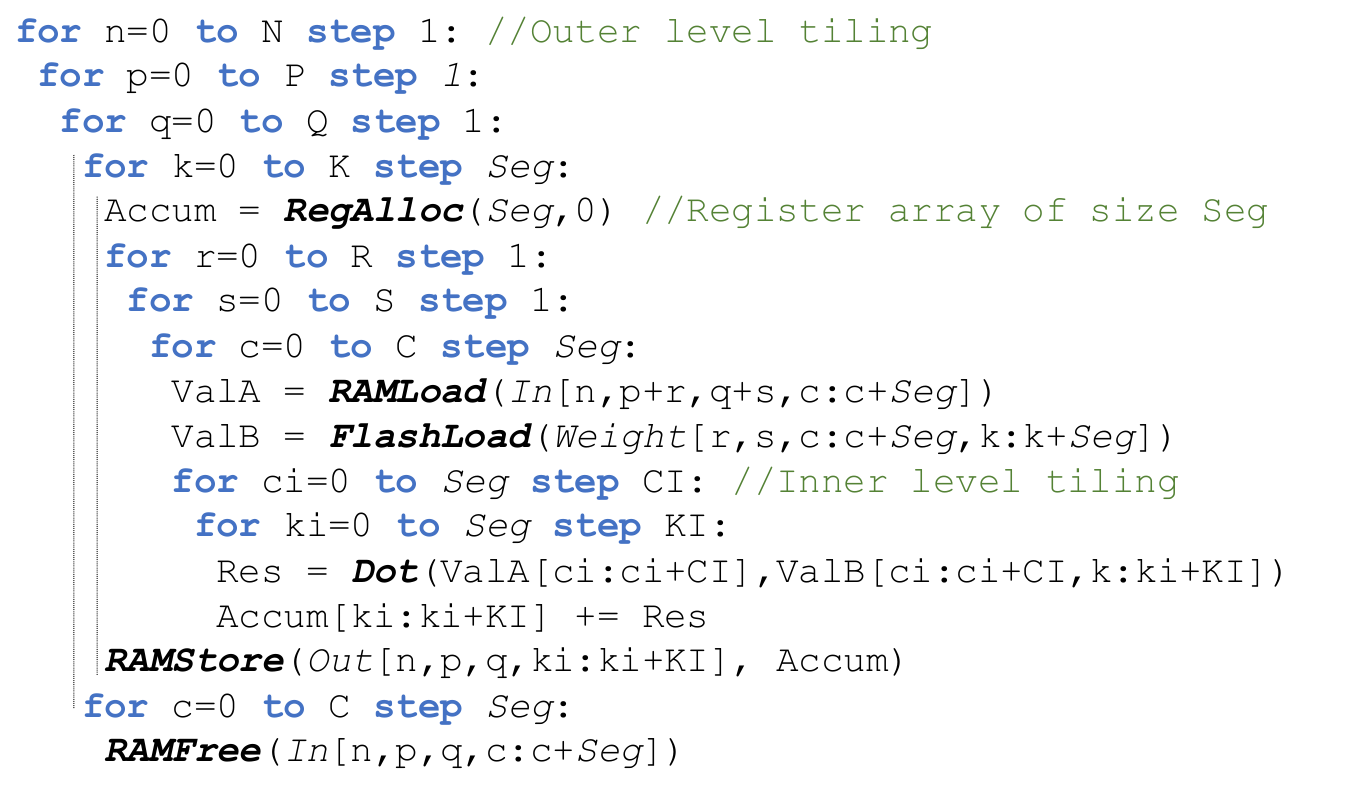}
\DeclareGraphicsExtensions.
\caption{Pseudo code for the kernel of 2D convolution layer
}
\label{fig:conv2d}
\end{figure}

Then, we show the pseudo code for 2D convolution in Figure~\ref{fig:conv2d}, which is a bit more complex than that of fully-connected layer.
The input tensor is $In[N,H,W,C]$, weight tensor is $Weight[R,S,C,K]$, and the output tensor is $Out[N,P,Q,K]$, where $P=H-R+1, Q=W-S+1$.
$N$ represents batch, $H, W$ represent image size, $C, K$ are input and output channels. $R, S$ are convolution window size. 
The two-level tiling sketch is similar to that of fully connected layer.
We also use the five functions mentioned above to implement this kernel.

\subsection{Kernel Design for Multiple Layers}

\begin{figure}[!t]
\centering
\includegraphics[width=3.3in]{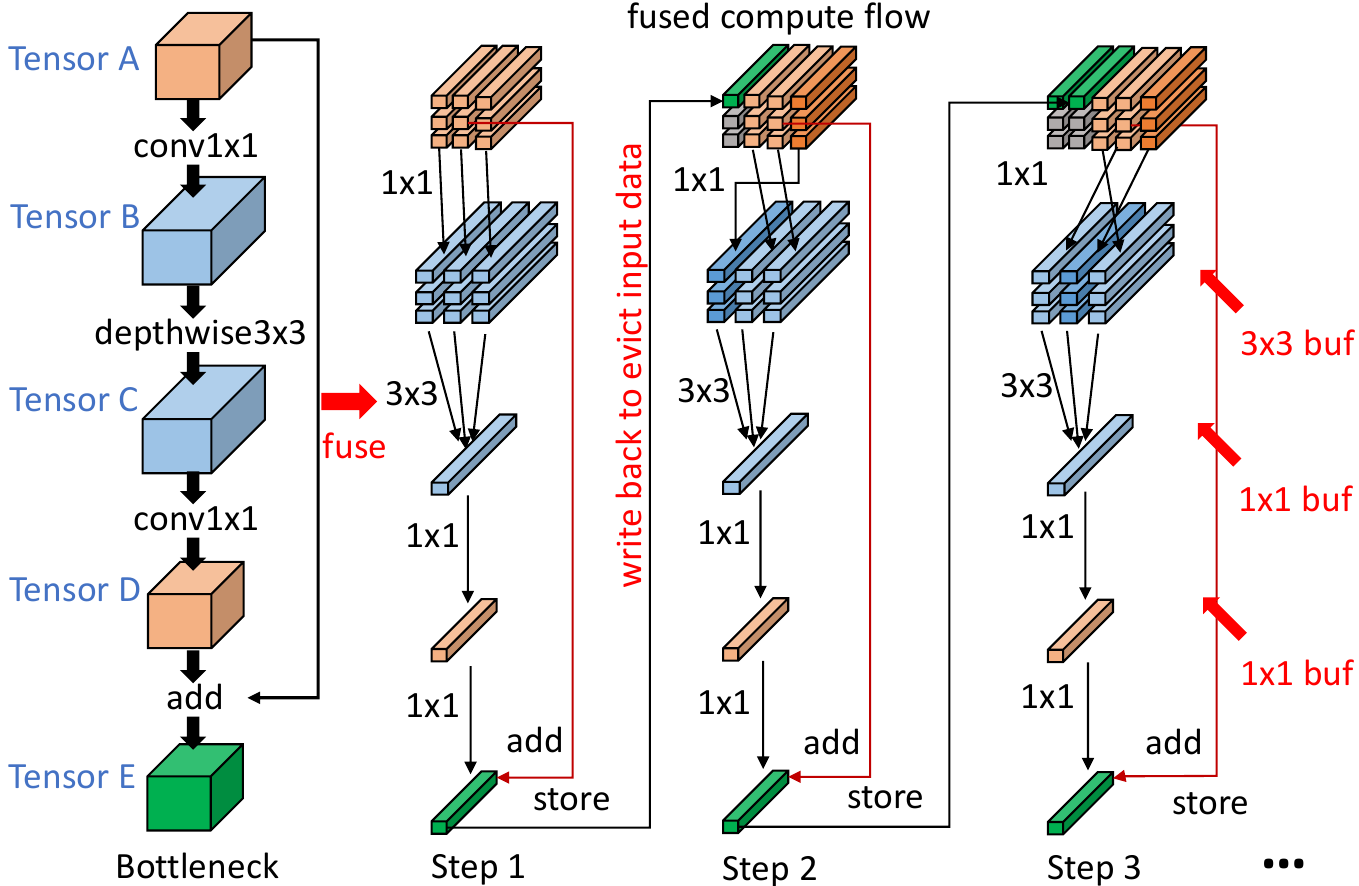}
\DeclareGraphicsExtensions.
\caption{Kernel design for inverted bottleneck module
}
\label{fig:sub-graph}
\end{figure}

Segment-level memory management is not restricted to a single layer but also applicable to multiple layers. 
The benefit of memory footprint reduction for a single layer is bounded by $50\%$ because we can ideally eliminate the memory overhead of either input or output tensor.
But considering multiple layers provides additional optimization possibility: we can eliminate the memory overhead of both input and output tensors for intermediate layers (layer fusion), enabling us to achieve better memory footprint reduction beyond $50\%$.
Designing a fused kernel for multiple layers is about deciding the scheduling of intermediate data (workspace) and the scheduling of input and output tensors.

Formally, this problem is a generalization of that of single layer (Section~\ref{sec:algorithm}).
A multi-layer module in DNN can be represented as a graph $G=(V, E)$, $V$ is a set of layers, $E$ is a set of edges between operators, representing the data producer-consumer relationships in the graph. 
For memory scheduling, the main constraint is to make sure that any output data write address of a consumer layer will never be used by a input data read access from its producer layer. This constraint should be conformed by all the operators in the graph. As a result, for any two layers $op_1, op_2 \in V$ that have producer-consumer relationship ($op_1$ is producer), for any iteration $\vec{j}$ from the iteration domain of $op_2$ and any iteration $\vec{i}$ from the iteration domain of $op_1$, if $\vec{j} \le \vec{i}$ (lexicographical order), then the following constraint $\mathcal{C}(op_1, op_2)$ holds

\begin{footnotesize}
\begin{equation*}
\begin{aligned}
& \mathcal{C}(op_1, op_2) :\ \  \forall{\vec{j} \le \vec{i}}, \ \vec{i} \in \text{domain of } op_1, \ \  \vec{j} \in \text{domain of } op_2\\
& \vec{L}_{In}^{op_1} (\mathbf{A}_{In}^{op_1} \vec{i} + \vec{V}_{In}^{op_1}) + b_{In}^{op_1} \ge \vec{L}_{Out}^{op_2} (\mathbf{A}_{Out}^{op_2} \vec{j} + \vec{V}_{Out}^{op_2}) + b_{Out}^{op_2}\\
\end{aligned}
\end{equation*}
\end{footnotesize}

And the optimization problem is defined as
\begin{footnotesize}
\begin{equation}
\label{eq:graph-formulation}
\begin{aligned}
& \text{min.} \ \ b_{In^*} - b_{Out^*}\\
& \text{s.t.} \ \ \mathcal{C}(op_1, op_2) \text{ holds, } \forall (op_1, op_2) \in E
\end{aligned}
\end{equation}
\end{footnotesize}where $In^*$ and $Out^*$ are input and output tensors for the whole graph.

The optimization problem is usually hard to be generally solved because there exist multiple valid solutions with non-trivial memory footprint and latency to tradeoff. But for a specific case, we can design an efficient kernel.
Considering that DNNs on MCUs are usually composed of variants of inverted bottleneck modules~\cite{mobilenetv2}, we show how to design efficient kernels for inverted bottleneck in this section.

Inverted bottleneck module is composed of four layers: two pointwise convolution layers, one depthwise convolution layer, and one add layer, as shown in Figure~\ref{fig:sub-graph} left part. We show all the five tensors in inverted bottleneck (weight tensors are not considered as they are in Flash).
For each step of computation, the kernel first loads $3\times 3$ segments from tensor A and produces $3\times 3$ segments for tensor B, which are stored in workspace; these $3\times 3$ segments are used to produce one segment of tensor C after depthwise convolution; then, the second pointwise convolution uses the segment of tensor C to produce a segment of tensorD, and the segment of tensor D is added with the corresponding segment of tensor A loaded previously, producing a segment of the final output tensor E.
The output segment for tensor E is updated back to the memory pool, which is handled by \ours{} memory management part.
The output segment can be placed at the memory address of a previously freed input segments, enabling memory footprint reduction just as what \ours{} does for single layer.
Overall, the kernel for inverted bottleneck needs additional 11 ($=3\times 3 + 1 + 1$) segments as workspace for intermediate data, which is a tiny overhead compared to the eliminated memory overhead of intermediate layers.

\subsection{Segment Size Selection}
Different segment size selection may affect the memory footprint and performance. For memory footprint, the smaller the segment size, the less the memory footprint is. This is because memory management is based on segments. If a segment contains only one element, then the achieved memory footprint is minimal. However, small segment size may cause high latency in execution because in our kernel design, we need to perform address modulo operations for data load and store. The smallest segment size (an element) requires modulo operations for each element loaded and stored, which will significantly reduce performance. As a result, we take a compromise. For fully connected layer, the segment size is the minimum of the row size of input and output tensor; for 2D convolution and inverted bottleneck module, the segment size is the minimum of the input and output channel size.

\section{\ours{} Compiler Support}
\label{sec:compiler}


We provide a Python programming interface for \ours{} to reduce the development difficulty. The Python code will be translated into intermediate representations (IR) and the IR is translated to low-level C++ code for MCU.
To achieve the best performance on MCU, we need to leverage the vector instruction set.
To expose the special instructions to Python programming interface, we provide one level of instruction wrapper called intrinsic (inlined functions).

\subsection{Vector Intrinsic Support}
We provide six types of intrinsic for programming: \textit{RegAlloc}, \textit{RAMLoad}, \textit{FlashLoad}, \textit{Dot}, \textit{RAMStore}, \textit{RAMFree}, and \textit{Broadcast}. We mainly explain three of them. The rest of them can be translated to standard C++ semantics in code generation.

\noindent
\textbf{Dot.} This intrinsic implements a fixed size matrix multiplication for problem size $2\times 2\times 16 $ with Int8 input data type and Int32 accumulator data type. It will be translated as a sequence of instructions using \textit{\_\_SADD16} and \textit{\_\_SMLAD} on ARM MCU.

\noindent
\textbf{RAMLoad.} This intrinsic implements vector load. This intrinsic will be translated into \textit{memcpy} operations on MCU.

\noindent
\textbf{Broadcast.} Broadcast intrinsic is useful for quantization operations on MCU. This intrinsic is translated into \textit{\_\_PKHBT} instruction on ARM MCU.

\subsection{Library Generation}
The generated code of \ours{} is compiled using ARM GCC toolchain for deployment.
To deploy the machine binary to target MCUs, we use ARM Mbed command line tools.
We pack the generated code of \ours{} for fully connected layers, 2D convolution layers, and inverted bottleneck modules into a light library for MCU. The library support dynamic input shapes so that the code size won't grow with input configurations.
\section{Evaluation}

\subsection{Evaluation Setup}
We use two platforms in evaluation: STM32-F411RE SoC (with Cortex-M4 processor, 128KB RAM) and STM32-F767ZI SoC (with Cortex-M7 processor, 512KB RAM). For single layer evaluation, our baseline is TinyEngine~\cite{mcunet, mcunetv2}.
The memory management strategies of TensorFlow Lite Micro~\cite{tflite-micro}, microTVM~\cite{tvm}, and CMSIS-NN~\cite{cmsis-nn} are similar to TinyEngine, but TinyEngine provides state-of-the-art latency as reported in MCUNet~\cite{mcunet, mcunetv2}, so we only compare to TinyEngine.
TinyEngine uses code generation with templates for convolutions and inverted bottleneck module. It supports inplace operations for depthwise convolution. But for other layers, it allocates different memory spaces for both inputs and outputs.
For multi-layer evaluation, we compare to TinyEngine and HMCOS~\cite{hierarchy}.
HMCOS is designed to schedule irregular graphs to reduce peak memory footprint. It doesn't support inplace operations.
In evaluation, we mainly report the RAM usage and energy consumption.

\begin{table}[t]
  \centering
  \footnotesize
  \caption{Configurations of \textit{inverted bottlenecks}.}
\begin{footnotesize}
  \begin{tabular}{c|c|c|c|c|c|c}
  \hline
    \textbf{Name} & \textbf{H/W} & \textbf{C\_in} & \textbf{C\_mid} & \textbf{C\_out} & \textbf{R/S} & \textbf{strides}\\
  \hline
  \multicolumn{7}{c}{\textbf{MCUNet-5fps-VWW}}\\
   \hline
    S1 & 20 & 16 & 48 & 16 & 3 & 1,1,1\\
  \hline
    S2 & 20 & 16 & 48 & 16 & 3 & 1,1,1\\
  \hline
    S3 & 10 & 24 & 144 & 16 & 3 & 1,1,1\\
  \hline
    S4 & 10 & 24 & 120 & 24 & 3 & 1,1,1\\
  \hline
    S5 & 5 & 40 & 240 & 40 & 3 & 1,1,1\\
  \hline
    S6 & 5 & 48 & 192 & 48 & 3 & 1,1,1\\
  \hline
    S7 & 3 & 96 & 480 & 96 & 3 & 1,1,1\\
  \hline
    S8 & 3 & 96 & 384 & 96 & 3 & 1,1,1\\
  \hline
    
  \multicolumn{7}{c}{\textbf{MCUNet-320KB-ImageNet}}\\
  \hline
    B1 & 176 & 3 & 16 & 8 & 3 & 2,1,1\\
  \hline
    B2 & 88 & 8 & 24 & 16 & 7 & 1,2,1\\
  \hline
    B3 & 44 & 16 & 80 & 16 & 3 & 1,1,1\\
  \hline
    B4 & 44 & 16 & 80 & 16 & 7 & 1,1,1\\
  \hline
    B5 & 44 & 16 & 64 & 24 & 5 & 1,1,1\\
  \hline
    B6 & 44 & 16 & 80 & 24 & 5 & 1,2,1\\
  \hline
    B7 & 22 & 24 & 120 & 24 & 5 & 1,1,1\\
  \hline
    B8 & 22 & 24 & 120 & 24 & 5 & 1,1,1\\
  \hline
    B9 & 22 & 24 & 120 & 40 & 3 & 1,2,1\\
  \hline
    B10 & 11 & 40 & 240 & 40 & 7 & 1,1,1\\
  \hline
    B11 & 11 & 40 & 160 & 40 & 5 & 1,1,1\\
  \hline
    B12 & 11 & 40 & 200 & 48 & 7 & 1,2,1\\
  \hline
    B13 & 11 & 48 & 240 & 48 & 7 & 1,1,1\\
  \hline
    B14 & 11 & 48 & 240 & 48 & 3 & 1,1,1\\
  \hline
    B15 & 11 & 48 & 288 & 96 & 3 & 1,2,1\\
  \hline
    B16 & 6 & 96 & 480 & 96 & 7 & 1,1,1\\
  \hline
    B17 & 6 & 96 & 384 & 96 & 3 & 1,1,1\\
  \hline
  \end{tabular}
  \label{table:inverted-bottlenect-config}
\end{footnotesize}
\end{table}

\begin{figure}[!t]
\centering
\includegraphics[width=3.3in]{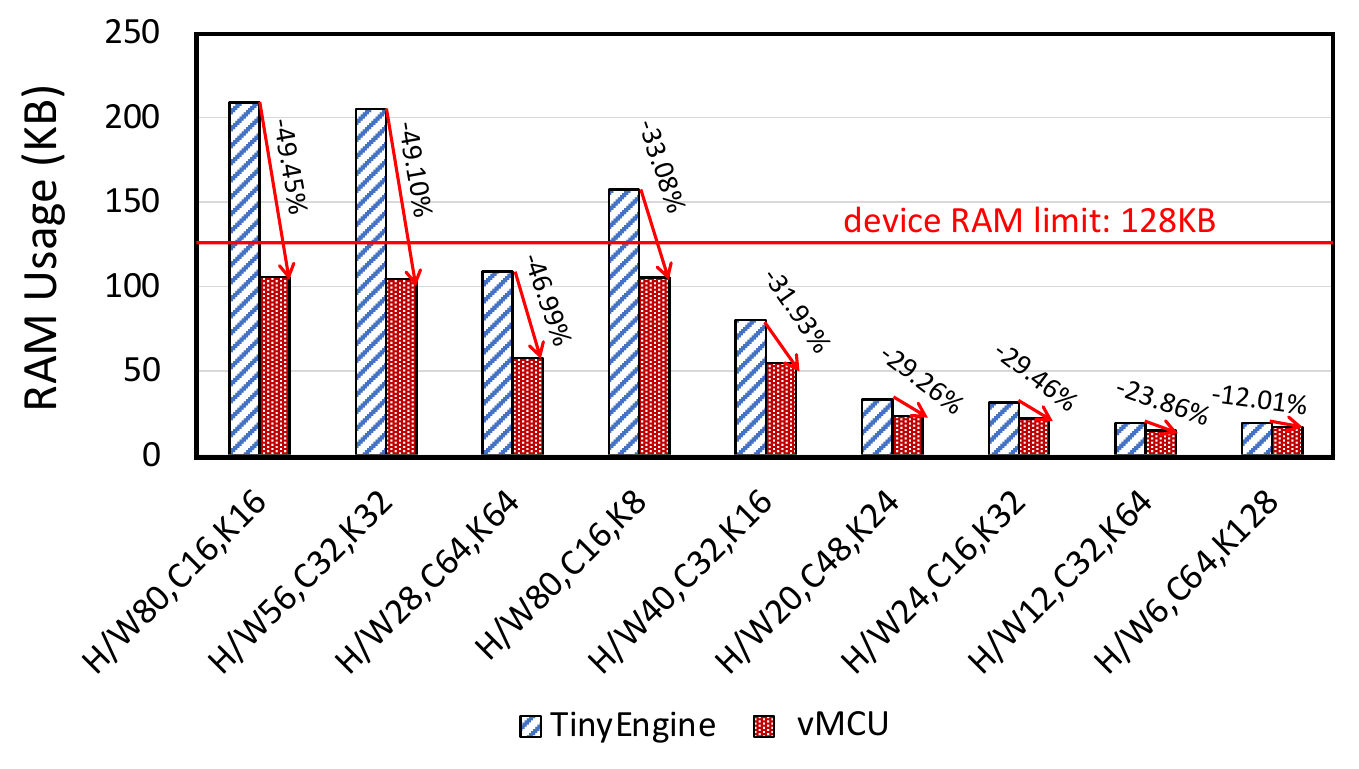}
\DeclareGraphicsExtensions.
\caption{RAM usage evaluation on STM32-F411RE.
}
\label{fig:single-op-results}
\end{figure}

\begin{figure}[!t]
\centering
\includegraphics[width=3.3in]{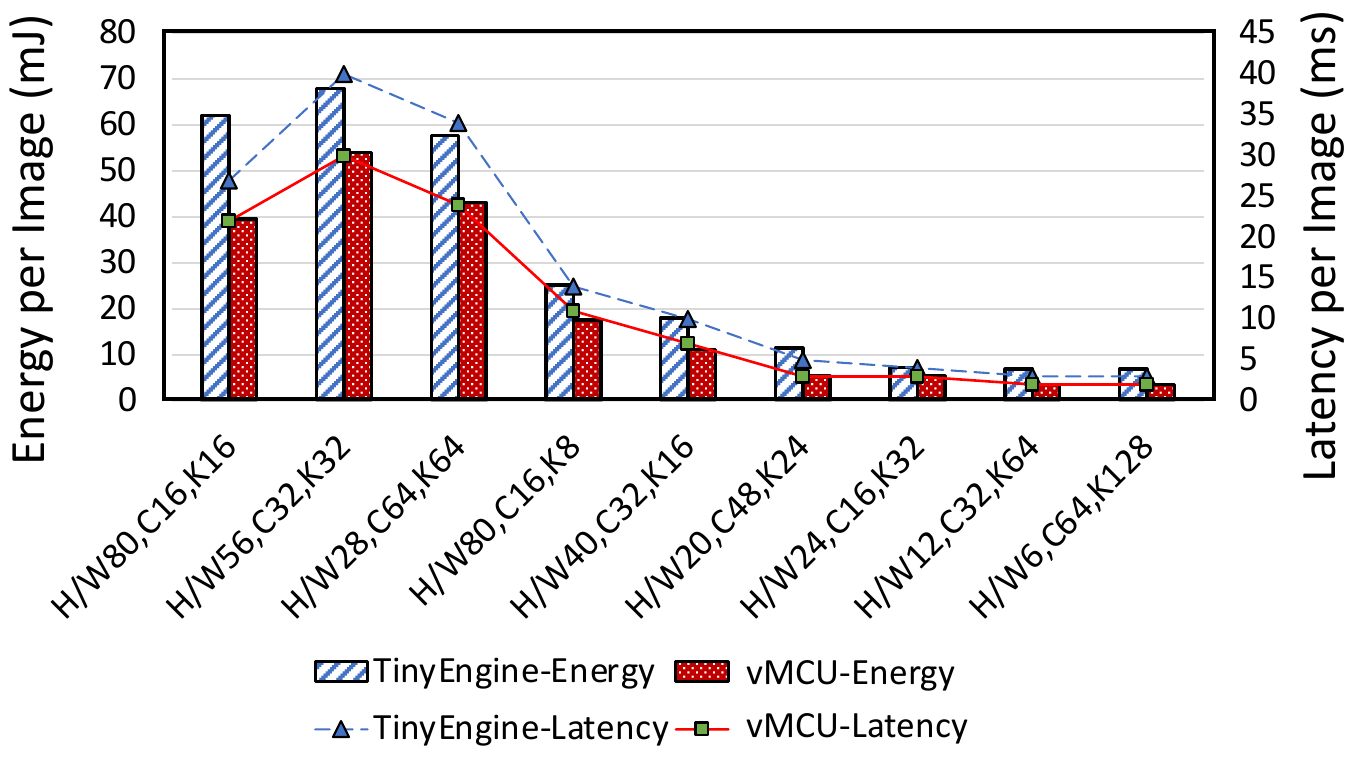}
\DeclareGraphicsExtensions.
\caption{Energy consumption and latency evaluation on STM32-F767ZI.
}
\label{fig:single-op-energy}
\end{figure}

\subsection{Single Layer Evaluation Results}

For single layer evaluation, we use pointwise convolution with nine different input shapes.  We use pointwise convolution because the current CNNs deployed on MCU mainly use pointwise convolution and depthwise convolution. For depthwise convolution, our memory management technique gives the same results compared to the inplace optimization in TinyEngine.

In Figure~\ref{fig:single-op-results}, we show the image size ($H/W$), input channel size ($C$), and output channel size ($K$) in the name of each case.  For example, $H/W80,C16,K16$ means that the convolution input image is $80\times 80$ with $16$ input channels, and the output channel is $16$. We use STM32-F411RE in evaluation, which only has 128KB RAM storage.
As the results show, \ours{} reduces memory footprint for all the nine cases compared to TinyEngine. The memory reduction ratio ranges from $12.01\%$ to $49.45\%$. For three cases (case 1, 2, 4), TinyEngine fails to run because it exceeds the RAM limit, while \ours{} successfully deploy all the convolutions to our target platform.
The memory reduction ratio for the first three cases are better than the others because these layers have larger activation size, which are common in the initial layers of a DNN graph (e.g., in ResNet~\cite{resnet} and MobileNet~\cite{mobilenetv2}). In addition, these layers' output activation sizes are the same as the input activation sizes, which makes the reduction ratio close to $50\%$. This means that we can eliminate about half of the memory consumption (the actual results are less than $50\%$ because we need extra memory space for output activation data as shown in Figure~\ref{fig:gemm-example}).
When the output activation size is different from that of input size, we can only eliminate the cost of the minimal one between them, resulting less memory reduction ratio  as shown in Figure~\ref{fig:single-op-results} case 4-9.
But overall, the memory usage of \ours{} is still less than TinyEngine.

Besides RAM usage, we also compare the energy consumption of \ours{} and TinyEngine. We measure the energy consumption of processing a single image for all the nine cases and show the results in Figure~\ref{fig:single-op-energy}. \ours{} reduces from $20.6\%$ to $53.0\%$ energy compared to TinyEngine. The energy consumption of MCU is highly related to the total number of memory accesses and execution latency.  TinyEngine consumes more energy mainly for two reasons. First, TinyEngine uses im2col algorithm to pre-process the input image of convolution. Although im2col is not necessary for pointwise convolution, TinyEngine doesn't bypass the pre-processing step, which results in more RAM accesses compared to \ours{}. Second, TinyEngine only unrolls loops to a predefined depth (e.g., 16), while \ours{} can fully unroll the innermost reduction loops to reduce processor pipeline stalls. 
We also plot the latency of each case in the secondary axis in Figure~\ref{fig:single-op-energy}. Overall, \ours{} can reduce the latency by from $18.5\%$ to $40.0\%$ compared to TinyEngine.

\begin{figure}[!t]
\centering
\includegraphics[width=3.3in]{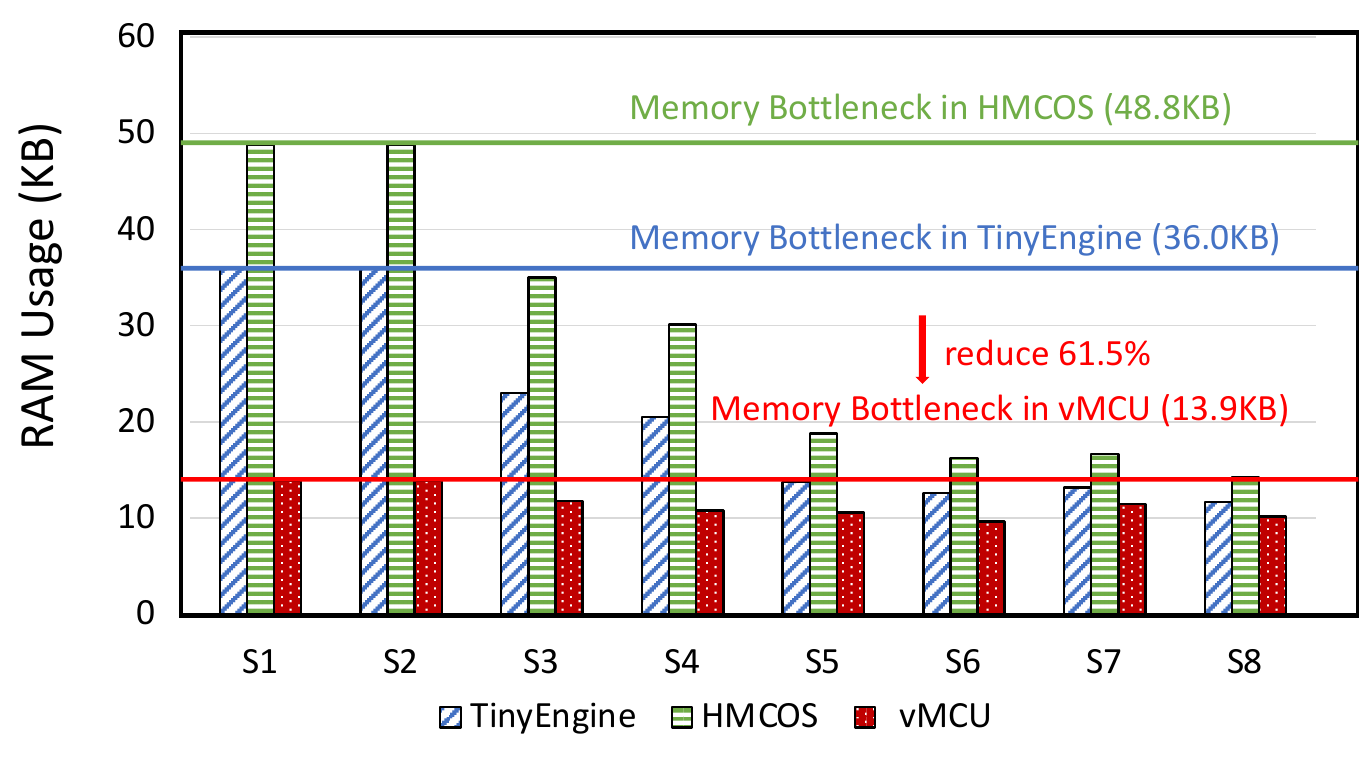}
\DeclareGraphicsExtensions.
\caption{RAM usage evaluation of inverted bottleneck for MCUNet-5fps-VWW on STM32-F411RE.
}
\label{fig:vww-results}
\end{figure}

\subsection{Multi-layer Evaluation Results}

For multi-layer evaluation, we use the inverted bottleneck modules from two real network: MCUNet-5fps-VWW and MCUNet-320KB-ImageNet~\cite{mcunet}.
The backbone of MCUNet-5fps-VWW is mainly composed of 8 inverted bottleneck modules, and
MCUNet-320KB-ImageNet is composed of 18 inverted bottleneck modules. 
We measure all the modules except for the last one for MCUNet-320KB-ImageNet because the last one uses a kernel size ($7\times 7$) that is larger than the image size ($6\times 6$), which is not suitable for fusion (but this module is not the memory bottleneck, so it doesn't affect the evaluation results). The configurations of all the modules are shown in Table~\ref{table:inverted-bottlenect-config}. \textbf{H/W} indicates the input images size, \textbf{C\_in} is the input image channel size, \textbf{C\_mid} is the output channel size of the intermediate depthwise convolution, \textbf{C\_out} is the output channel size of the module, \textbf{R/S} is the kernel size of the depthwise convolution, and \textbf{strides} are the strides of the three convolutions in the module.

The evaluation results of MCUNet-5fps-VWW are shown in Figure~\ref{fig:vww-results}.
This network is small and can be deployed to STM32-F411RE. Although TinyEngine, HMCOS, and \ours{} all successfully deploy the network to target MCU, \ours{}'s memory usage is the minimal.
Compared to HMCOS, \ours{} reduces RAM usage by $28.8\% - 71.6\%$; compared to TinyEngine, \ours{} reduces RAM usage by $13.0\% - 61.5\%$.
The memory bottleneck of this network is the first module. For this module, compared to HMCOS, \ours{} reduces the memory bottleneck by $71.6\%$; compared to TinyEngine, \ours{} reduces the memory bottleneck by $61.5\%$.
We also evaluate the latency of each inverted bottleneck module for MCUNet-5fps-VWW. The results are listed in Table~\ref{table:inverted-bottlenect-latency}. Overall, the latency of \ours{} is comparable to that of TinyEngine ($1.03\times$).

\begin{table}[t]
  \centering
  \footnotesize
  \caption{Latency results of \textit{inverted bottlenecks} in MCUNet-5fps-VWW.}
\begin{footnotesize}
  \begin{tabular}{c|c|c|c}
  \hline
    \multirow{2}{*}{\textbf{Name}} & \textbf{Latency} & \textbf{Throughput} & \textbf{TinyEngine's}\\
    & \textbf{(ms)} & \textbf{(image/s)} & \textbf{Latency(ms)}\\
   \hline
    S1 & 37 & 27 & 37\\
   \hline
   S2 & 37 & 27 & 37\\
   \hline
   S3 & 33 & 30 & 35\\
   \hline
   S4 & 28 & 35 & 29\\
   \hline
   S5 & 22 & 45 & 24\\
   \hline
   S6 & 20 & 50 & 19\\
   \hline
   S7 & 34 & 29 & 36\\
   \hline
   S8 & 27 & 37 & 28\\
   \hline

  \hline
  \end{tabular}
  \label{table:inverted-bottlenect-latency}
\end{footnotesize}
\end{table}

\begin{figure}[!t]
\centering
\includegraphics[width=3.3in]{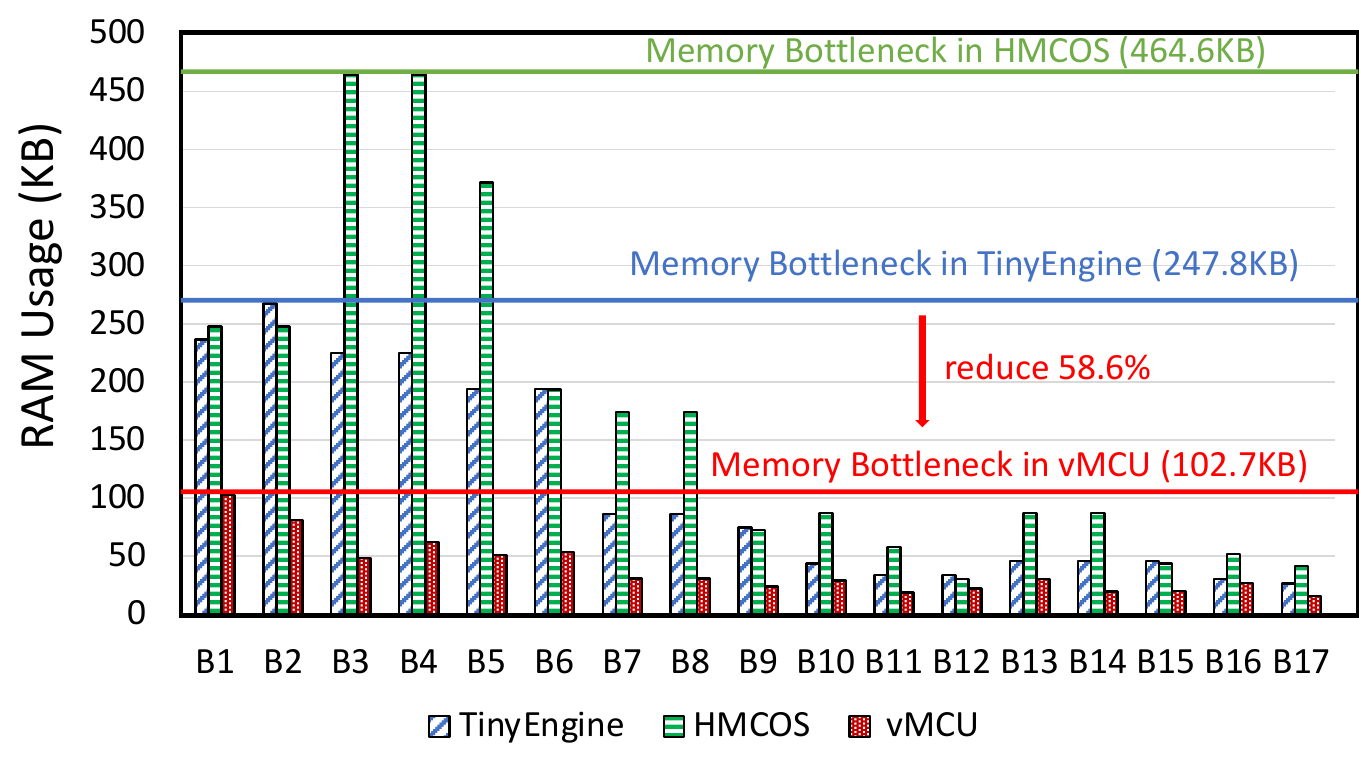}
\DeclareGraphicsExtensions.
\caption{RAM usage evaluation of \textit{Inverted Bottlenecks} for MCUNet-320KB-ImageNet on STM32-F767ZI.
}
\label{fig:bottleneck-results}
\end{figure}

The evaluation results for MCUNet-320KB-ImageNet are shown in Figure~\ref{fig:bottleneck-results}. 
MCUNet-320KB-ImageNet is larger than MCUNet-5fps-VWW and can't be deployed to STM32-F411RE by HMCOS or TinyEngine.
By contrast, \ours{} can reduce the RAM usage by $26.5\% - 89.6\%$ compared to HMCOS. Compared to TinyEngine, \ours{} can reduce $11.2\% - 78.5\%$ RAM usage. HMCOS fails to reduce memory space for such linear structure DNNs. Compared to HMCOS, TinyEngine can use inplace depthwise convolution to reduce memory footprint. But the memory footprint of pointwise convolutions is not reduced. When deploying the whole network, the memory bottleneck of HMCOS is 464.6KB (B3); the bottleneck of TinyEngine is 247.8KB (B2); while the bottleneck of \ours{} is 102.7KB (B1). We reduce the bottleneck by $58.6\%$ compared to TinyEngine. As a result, we can successfully deploy this network to a more resource-constrained platform STM32-F411RE (128KB RAM).

\begin{figure}[!t]
\centering
\includegraphics[width=3.3in]{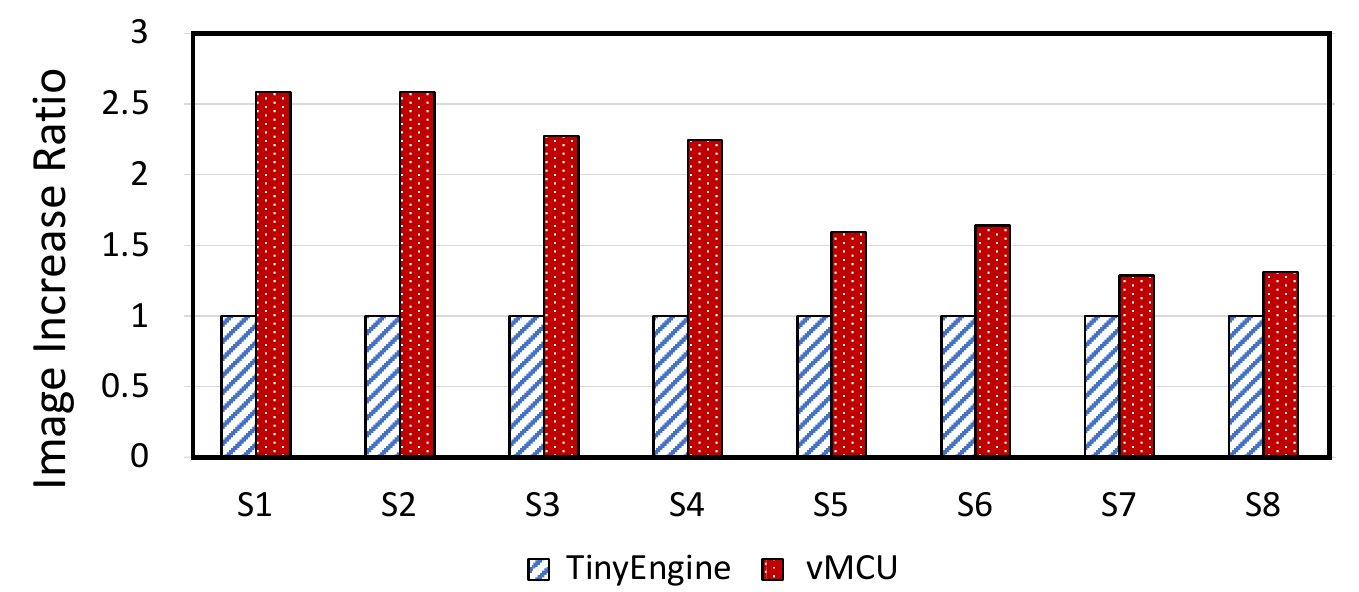}
\DeclareGraphicsExtensions.
\caption{The increase ratio for image size (both height and width) of \ours{} compared to TinyEngine when using the same amount of RAM.
}
\label{fig:increase-image}
\end{figure}

\subsection{Impact on Accuracy and NAS}
\begin{figure}[!t]
\centering
\includegraphics[width=3.3in]{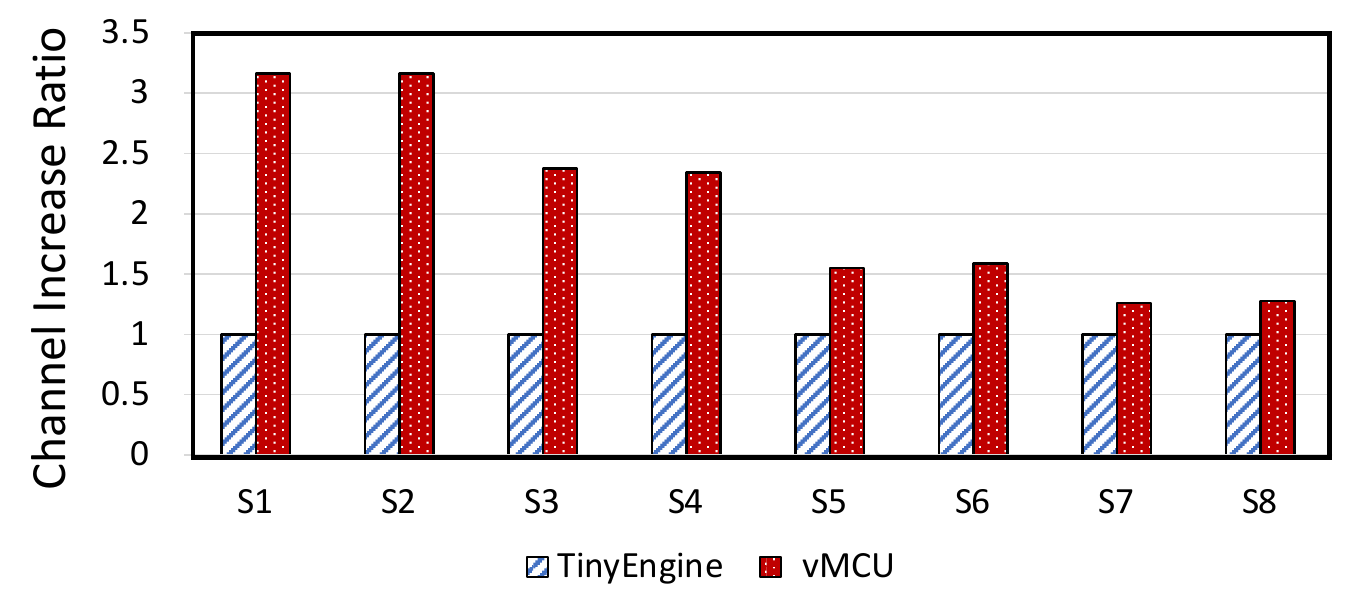}
\DeclareGraphicsExtensions.
\caption{The increase ratio for channel size (both input and output channel) of \ours{} compared to TinyEngine when using the same amount of RAM.
}
\label{fig:increase-channel}
\end{figure}

The optimizations in \ours{} do not change the original correctness of the computation and thus the same DNNs can be deployed to MCUs without any accuracy loss.
Moreover, \ours{} enables a larger optimization space for NAS because the memory usage of each layer and multi-layer module can be significantly reduced and the memory constraints in the NAS process~\cite{mcunet} can be relaxed. For example, when using exactly the same amount of memory resources as TinyEngine, \ours{} allows the network MCUNet-5fps-VWW to increase its image size (both height and width) by from $1.29\times$ to $2.58\times$ or increase its channel size (both input channel and output channel) by from $1.26\times$ to $3.17\times$ as shown in Figure~\ref{fig:increase-image} and Figure~\ref{fig:increase-channel}, which means that more compute operations (OPs) are allowed in deployment and the network accuracy can be potentially increased by retraining.

\section{Related Work}

\subsection{Hand-tuned Libraries for MCU}
Hand-tuned libraries for MCU focus on optimizing the latency of single operators through tiling and instruction selection.
CMSIS-NN~\cite{cmsis-nn} implements various operators including matrix multiplication, convolution, and softmax. It optimizes these operators by using small tiling factors (e.g., compute 2 rows for convolution at a time) and use the SIMD instructions provided by MCU.
CMix-NN~\cite{cmix-nn} implements low-precision operators (2bit, 4bit, 8bit) for matrix multiplication and convolution.
Deep learning frameworks such as TensorFlow Lite Micro~\cite{tflite-micro} rely on these libraries for deployment.
These libraries focus on latency optimization or bit-width optimization without consideration for memory footprint reduction.

\subsection{Machine Learning Compiler}
Machine learning compilers such as Halide~\cite{halide}, TVM~\cite{tvm}, MLIR~\cite{mlir}, FlexTensor~\cite{flextensor}, Ansor~\cite{ansor}, AMOS~\cite{amos}, TensorIR~\cite{tensorir}, and Chimera~\cite{chimera} provide the abilities to generate high-performance code on a wide range of hardware backends. Among them, compilers for MCUs focus on both latency optimization and memory reduction. TinyEngine~\cite{mcunet, mcunetv2} provides code generation support for CNNs and support in-place memory optimization at tensor level. \textsc{Magis}~\cite{magis} uses fission optimization to reduce memory footprint for DNN graphs. Compared to them, \ours{} uses segment-level memory management to deploy DNNs on MCU and uses high-level programming model help operator development.

\subsection{Memory Optimization with Model Redesign}
NAS~\cite{proxyless-nas} is used to find an optimized model architecture that gives the best accuracy, latency, and peak memory footprint for MCU.
MCUNet~\cite{mcunet} uses NAS to design CNNs with extremely small memory requirements on MCU. The result networks only need hundreds of kilobytes RAM resources for execution.
MCUNet-V2~\cite{mcunetv2} splits the initial convolution layers of CNNs into partial convolutions to further reduce memory requirements at the cost of re-computation.
To alleviate the overhead of re-computation, NAS is used to redesign convolution kernel sizes.
Besides NAS, TREC-based~\cite{trec} methods retrains the DNNs to eliminate redundant computation from input data.
These approaches need to retrain or fine-tune existing networks.
Compared to them, \ours{} doesn't need any modification of the original DNN model parameters.

\subsection{Tensor-level Memory Management}
Tensor-level memory management reduces the memory footprint by overlapping different tensors in memory space.
Serenity~\cite{chaos} searches the optimal execution order through dynamic programming and overlaps tensor with different lifetime.
HMCOS~\cite{hierarchy} improves the searching method by first finding the sub-graph that is memory bottleneck and then optimizing only the sub-graph instead of the whole graph.
Tensor compilers such as Relay~\cite{relay}, nGraph~\cite{ngraph}, and TASO~\cite{taso} mainly optimize the execution order using fusion. They lack optimizations for memory footprint reduction.
Moreover, the scheduling techniques are only effective for irregular network structure. For linear structure, there is little or no benefit from scheduling.
TinyEngine~\cite{mcunet, mcunetv2} only overlaps input and output tensors for depthwise convolution and can't apply overlapping for fully connected layers or 2D convolution layers.
COMB~\cite{comb} uses genetic algorithms to search for optimized schedules for DNNs on heterogeneous SoC. It optimizes data transfer overhead at tensor-level but doesn't reduce memory footprint and thus can't save memory for MCUs.
Compared to them, \ours{} can reduce the memory footprint of networks with linear structure by segment-level memory management.

\section{Conclusion}

Deploying DNN models to IoT devices that are based on microcontrollers (MCU) is becoming prevalent. To fully utilize the constrained storage resource of MCUs, we propose \ours, which uses segement-level memory management techniques to reduce memory footprint for both single layer and multi-layer scenarios. 
The evaluation shows that our technique can reduce from $12.0\%$ to $49.5\%$ RAM usage and from $20.6\%$ to $53.0\%$ energy consumption compared to state-of-the-art work for single layer; for end-to-end graph evaluation, our technique can reduce memory bottleneck by $61.5\%$.

\section*{Acknowledgements}
We thank all the anonymous reviewers for their insightful suggestions. This work is supported in part by the National Natural Science Foundation of China (NSFC) under grant No.U21B2017.




\bibliography{ref}
\bibliographystyle{mlsys2024}



\end{document}